# AlphaFold two years on: validation and impact

Oleg Kovalevskiy[1], Juan Mateos-Garcia[1], Kathryn Tunyasuvunakool[1]
[1]Google DeepMind, London, UK

## Abstract

*Two years on from the initial release of AlphaFold2 we have seen its widespread adoption as a structure prediction tool. Here we discuss some of the latest work based on AlphaFold2, with a particular focus on its use within the structural biology community. This encompasses use cases like speeding up structure determination itself, enabling new computational studies, and building new tools and workflows. We also look at the ongoing validation of AlphaFold2, as its predictions continue to be compared against large numbers of experimental structures to further delineate the model's capabilities and limitations.*

## Introduction

The first experimental protein structures were determined in the 1950s using X-ray crystallography, proving that protein chains fold into well-defined 3D shapes (1). Anfinsen went on to assert that each sequence of amino acids adopts a specific 3D structure (2). This raised an important question - can we predict a protein's structure given only its amino acid sequence?

Even in the early days of structural biology, when few protein structures were available, it became clear that proteins with similar sequences adopt similar shapes. This quickly led to the idea of homology modelling - predicting a structure based on its sequence similarity to known structures, see e.g. (3). As more experimental data accumulated, the Protein Data Bank (PDB) was established in the 1970s (4). This was crucial for the field, making possible open sharing of structural data, facilitating its analysis, and laying the foundation for all future structure prediction efforts.

In 1994, the Critical Assessment of Methods for Protein Structure Prediction (CASP) was started to encourage the development of more accurate prediction methods (5). CASP consists of blind prediction challenges: participants predict non-trivial protein structures for which experimental results have not yet been made public. With the growth in computing power and in the PDB during the late 90s and 2000s,

*Corresponding author: okovalevskiy@deepmind.com*

novel computational methods flourished. Fragment-based methods used short protein fragments extracted from experimental structures as building blocks to construct a prediction (6). More recently, methods incorporating evolutionary information and contact prediction showed great promise (7–9). However, predictions rarely met the bar for near-experimental quality (a GDT_TS score > 90) before CASP14, when the machine learning system AlphaFold2 (referred in this paper as simply AlphaFold) achieved this level of accuracy on the majority of CASP targets (10, 11).

Since AlphaFold's release in 2021 it has seen rapid and widespread adoption. Based on data from OpenAlex, the paper describing the method has now received over 10,000 citations (10). Originated as a repository of 360,000 predicted protein structures from 21 organisms including humans, the AlphaFold Database has since grown exponentially, encompassing a staggering collection of 214,000,000 structures in 2022. The AlphaFold Database has had 1.34 million unique visitors from more than 190 countries, and the whole archive has been downloaded over 18,000 times (12–14). Finally, our own analysis of PDB depositions up to January 2023 found ~850 entries associated with a paper that uses AlphaFold (of which over 60% were cryo EM structures); evidence of the method's widespread use among experimental structural biologists. Here, we review how AlphaFold is being used today, with a particular focus on that community. We also discuss recent work on evaluating AlphaFold, by comparing its predictions against experimental results on a larger scale. Given the rapid pace of the field, we don't expect this review to be complete.

**AlphaFold in structural biology**

*Impact on experimental structure determination*
A major outcome of improvements in structure prediction has been to accelerate the work of experimental structural biologists, by simplifying certain steps in their workflow.

This first became apparent in X-ray crystallography. To determine a structure by X-ray crystallography or microcrystal electron diffraction, it is necessary to reconstruct the phase information lost during the diffraction experiment. Molecular replacement is a technique to do this that requires no additional experimental work, but it does rely on having a search model that closely resembles the target structure. There have now been numerous reports of successful molecular replacement using AlphaFold predictions (15–18), including challenging cases where all search models derived from PDB had failed (19, 20), where the target had a novel fold (21), or was a *de novo* design (22). In fact, work by Terwilliger et al. (discussed later) suggests that a high percentage of structures can now be phased largely automatically using AlphaFold (23).

The community has developed a variety of tools to support this workflow, making a further valuable contribution to accelerating the process. Both major software suites for macromolecular crystallography, CCP4 (24, 25) and PHENIX (26), now include import procedures that convert AlphaFold's pLDDT[1] confidence metric into an estimated B-factor and remove low-confidence regions. CCP4Cloud, a cloud-based environment for crystallographic computations (27), offers online access to AlphaFold modelling (25). Widely used automatic tools like MRBUMP (28) and MRPARSE (29) can now search for templates and fetch predictions from AlphaFold Database with minimal user intervention. In some cases it is better to split a prediction into smaller regions before attempting molecular replacement. Software like Slice'n'Dice in CCP4 (30) and PHENIX's process_predicted_model (31) can split an AlphaFold prediction into domains based on its PAE plot[2] or on spatial clustering, while ARCIMBOLDO (32), an *ab initio* phasing tool, can extract fragments from AlphaFold models (33). The Low Resolution Structure Refinement pipeline (LORESTR) has also been updated to automatically fetch models from AlphaFold Database and use them for restraints generation (25, 34).

AlphaFold has also had a substantial impact on structure determination by cryo EM. Since the "Resolution Revolution" (35) it is relatively common to obtain detailed electron density maps with resolution better than 3.5 Å. Still, many reconstructions suffer from data collection problems leading to lower resolution in some regions. Cryo ET and subtomogram averaging are also now used to visualise large assemblies or parts of whole cells, and can yield lower resolution data. Combining cryo EM with AlphaFold predictions can give the best of both worlds, with the experimental data serving to validate the prediction and reveal domain arrangements, while the prediction provides fine atomic details.

A pioneering example of this integrative approach was work on the nuclear pore complex, which fit AlphaFold models for individual proteins and small subcomplexes into electron density maps with resolution 12 - 23 Å, reconstituting the majority of this enormous ~120 MDa assembly (36, 37). Since then we have seen numerous other integrative cryo EM examples, elucidating the structures of the intraflagellar train (38–40); the augmin complex (41, 42); components of the yeast small subunit processome (43) and components of the eukaryotic lipid transport machinery (44). In the case of Retriever (part of a 0.5 MDa endosomal trafficking complex) particles suffered from preferred orientations, leading to one direction being poorly resolved with an overestimated overall resolution of 4.3 Å. However, the close agreement with AlphaFold's prediction made it easy to fit the model into experimental maps, and the authors went on to reconstruct the whole Commander complex, combining information from experimental structures and predictions (45).

---

[1] pLDDT (predicted Local Distance Difference Test) is a per-residue score between 0 and 100 that reflects the AlphaFold's confidence in the local structure of a domain.
[2] PAE (Predicted Aligned Error) is an AlphaFold confidence metric reported for each pair of residues, and reflects confidence in their relative positions. A low PAE implies high confidence.

Recognising the utility of this approach, some of the major model building and fitting programs used in cryo EM have added support for AlphaFold predictions. COOT (46) can import predictions from AlphaFold Database, while ChimeraX (47) includes an option to generate new predictions in ColabFold (48). Again the community has built on AlphaFold to produce useful automated workflows. For example, an iterative procedure for model building has been developed that begins by fitting an initial AlphaFold prediction into the experimental density using PHENIX tooling (49). In subsequent iterations, the latest fitted structure is provided to AlphaFold as a template, producing a prediction that more closely matches the density. This iterative procedure improves the resulting structures beyond simple rebuilding against experimental data. Another automated solution uses a deep-learning based quality score (DAQ) to identify low-quality regions, then rebuilds these in a targeted fashion with AlphaFold (50). Interestingly, an analysis using the ML-based validation tool *checkMySequence* has highlighted at least one example where a deposited cryo EM structure appears to suffer from a register shift, while a rebuilt model guided by an AlphaFold prediction is in good agreement with the experimental density map (51). Another ML tool *conkit-validate* specifically uses AlphaFold predictions to derive inter-residue contacts and distances for identification of register shifts (52).

A particularly novel use of AlphaFold in cryo EM has been identifying unknown densities via structural search. In one example, researchers were working to solve the structure of the mycobacterial lipid transporter Mce1 (53). Their density maps revealed a previously unknown subunit of the complex, in sufficient detail to build a poly-alanine model of the protein. They were then able to perform a structural search of the model against a large number of predictions in AlphaFold Database, which returned a hit for MSMEG_3032 / LucB. The assignment was subsequently validated by checking that LucB and the rest of the Mce1 system co-purify. This method in particular is only possible thanks to the availability of large prediction databases.

*Predicting protein-protein interactions*

Although AlphaFold initially was not trained to predict protein-protein complexes, it became apparent that even a monomer version of AlphaFold is capable of predicting them (54). A specially trained AlphaFold-Multimer was released later in 2021, facilitating the discovery and characterisation of new protein-protein (including protein-peptide) interactions (55). Computational methods are useful in this context, as they can scale to screening millions of protein pairs. One of the first examples was work by Humphreys et al., which used a combination of RoseTTAFold (56) and AlphaFold to screen 8.3 million protein pairs from *Saccharomyces cerevisiae (54)*. Searching for complexes that might be broadly conserved across eukaryotes, they identified 1,505 novel interactions, and proposed predicted structures for 912 assemblies. Other large scale interaction prediction efforts have explored the human proteome (57) and the proteome of *Bacillus subtilis (58),* using experimental data and prior knowledge to narrow down the set of protein pairs to process.

More recent work has used AlphaFold-Multimer to better understand specific biological pathways on a mechanistic level (55). For example, Gu et al. had identified the largely uncharacterized protein midnolin as a novel mediator of proteasomal degradation, involved in regulating levels of transcription factors like EGR1, FosB and c-Fos (59). They used AlphaFold-Multimer to predict the structure of midnolin in complex with several of its substrates, including IRF4. The complex prediction suggested a mechanism of action in which two Catch domains in midnolin come together to capture a β-strand portion of the substrate. This hypothesis was tested for several midnolin substrates, either by introducing targeted mutations into the predicted β-strand region or deleting it altogether, after which the interaction with midnolin no longer occurred.

In a separate example, Lim et al. were studying the protein DONSON, which is necessary for the assembly of CMG helicase in vertebrates (60). However, exactly how DONSON mediated helicase assembly was unclear. The authors used AlphaFold-Multimer to screen 70 core DNA replication factors for possible interactions with DONSON. Based on the most confidently-predicted complexes, they were able to build up a structural model of a pre-Loading Complex, in which DONSON interacts with GINS, TOPBP1, and Pol ε. Experimental evidence for the model was subsequently obtained from coimmunoprecipitation and site directed mutagenesis. Analogously, Sifri et al. investigated a system for DNA double-strand break repair, employing AlphaFold-Multimer to predict all possible pairwise protein combinations within the 53BP1-RIF1-shieldin-CST pathway. Their analysis revealed a novel binding interface between RIF1 and SHLD3 and provided structural information for seven previously characterised interactions; these findings were subsequently confirmed experimentally (61). These examples illustrate how multimeric structure prediction can shed light on protein protein interactions, both through large scale screens and more targeted structure modelling.

Today, software is available that aims to simplify and accelerate interaction screening with AlphaFold-Multimer, notably AlphaPulldown (62). Besides large scale screening, other supported use cases include locating the binding interface between two proteins by screening pairs of sequence fragments, and identifying which subunits of a complex are in direct contact via all-to-all screening. Given a list of pairwise subunit interactions, other tools like MoLPC can attempt to build out a model of the full complex (63). MoLPC uses Monte Carlo tree search to explore possible orders in which to assemble the chains, stopping when there are too many clashes and scoring each output to identify the most promising assembly. The latest update of AlphaFold-Multimer also supports higher residue and chain limits, meaning that complexes of up to 20 chains may now be predicted directly.

### *Use in protein design*

While AlphaFold is not intended as a protein design model, it has been used by the community as a component in design pipelines. For example, Wicky et al. used AlphaFold to generate novel "hallucinated" complexes, by beginning with a random sequence and number of copies, then performing Monte Carlo search using AlphaFold confidence plus a cyclic symmetry metric as the loss function (64). This generated topologically diverse 1- to 7-mers that were subsequently used as targets for design with ProteinMPNN (65). A more recent investigation has shown that both AlphaFold and RosettaFold are useful for filtering protein designs, with the inclusion of a structure prediction step increasing success rates significantly over a purely energy-based pipeline (66). After exploring several filtering approaches, the authors used AlphaFold average interchain PAE < 10 as a selection criterion in their prospective analysis, finding that this yielded a higher success rate for binder designs by between 8 and 30 fold.

Another way AlphaFold has been used in design is simply to generate a starting prediction for a protein with no experimental structure, which can then be used to guide design efforts and suggest which domains or residues to edit. For example, in recent work by Kreitz et al. the authors aimed to re-engineer a contractile injection system from the bacterium *Photorhabdus asymbiotica* so that it would target human cells. As no experimental structure was available, they modelled the trimeric distal tip protein with AlphaFold, revealing a globular domain that appeared to be responsible for target recognition. By replacing this domain with alternative binding proteins, they were able to alter the injection system to target human and mouse cells, demonstrating its potential as a delivery system for therapeutics (67).

*Enabling new computational work*

A key advantage of computational methods is their ability to scale. Experimental structural biology may take months to solve one protein structure, and so can't keep pace with the rapid accumulation of known protein sequences. Structure prediction tools can keep up with modern sequencing, and enable the construction of extremely large prediction databases. Examples include AlphaFold Database (12, 13) , which now covers over 200M UniProt sequences (68), and ESM Metagenomic Atlas (69), which covers 600M metagenomic sequences. The community has further enriched these prediction databases with information from other sources. For example, AlphaFill adds ligands to predictions by "transplanting" them in from similar PDB structures (70), while TmAlphaFold and AFTM use software to add predicted membrane planes (71)(72).

The availability of large structure databases has spurred on the development of efficient algorithms like FoldSeek (73) and FoldSeek cluster (74), which can group structurally similar entries or identify proteins similar to a query structure. We have already mentioned how structural search can be used to identify unknown densities in cryo-EM maps. It can also be applied in other situations where researchers would previously have relied on sequence homology, e.g. for functional annotation (75), or for identifying parasite proteins that are molecular mimics of a host protein (76).

Large structure prediction databases also contribute to the effort to catalogue protein folds. CATH, a hierarchy-based structural classification of protein domains, now incorporates AlphaFold predictions from 21 model organisms. Analysis of ~370,000 predicted domains, filtered based on confidence and geometrical quality, assigned 92% of them to existing CATH superfamilies. Nevertheless, AlphaFold predictions contained considerable structural novelty: 25 new superfamilies, and a 36% increase in the number of unique 'global' folds (77). More recent analyses of the full AlphaFold Structure Database have looked at the properties of its structural clusters (74), and its potential for use in function annotation (78).

Other computational work builds on AlphaFold by leveraging its confidence metrics rather than its ability to scale. Two particularly interesting examples explore AlphaFold's ability to rank its own predictions. Firstly, a method called AFSample has been developed which generates and ranks ~5,000 AlphaFold-Multimer predictions for any given input (79). Diversity is boosted by enabling dropout and by varying a range of settings (e.g. whether templates are used, and the number of recycling iterations). AFSample demonstrates that ranking large numbers of diverse predictions using a pTM-based score[3] often succeeds in picking out higher accuracy models, and the method ranked top in the protein assembly category of CASP15, with a +0.13 higher DockQ score than default AlphaFold-Multimer.

Meanwhile, Roney and Ovchinnikov have investigated whether AlphaFold can rank predictions made in the absence of any evolutionary information from a multiple sequence alignment (80). They generated multiple predictions for a given target structure, each using a different "decoy" structure as the input template. A "composite confidence score" based on several AlphaFold outputs was able to rank the resulting predictions with a mean Spearman's coefficient of 0.925, identifying the decoys closest to the target. Composite confidence outperformed ROSETTA's energy function (81) and DeepAccNet (82) at this task. The authors proposed based on the results that AlphaFold has "learned an energy function that can assess sequence-structure agreement, but needs coevolution data or templates to help search for optimal structures".

This observation resonates with another finding: AlphaFold's low confidence scores (low pLDDT and high predicted aligned error) correlate strongly with intrinsic protein disorder, making it a state-of-the-art predictor of protein disordered regions. This was initially observed during large-scale protein structure modelling for the AlphaFold Database (13) and later confirmed by independent researchers (17, 83, 84).

Although AlphaFold was originally designed to predict only one conformation for a particular sequence of amino acids, it seems that it is often possible to induce AlphaFold into generating alternative structural states of a protein. One way of doing this is subsampling or clustering of MSA based on sequence similarity (85)(86). Another strategy is combining shallow MSA with a template corresponding to the specific state (87)(88). If the MSA signal is strong, AlphaFold tends to ignore structural information from the template; artificial weakening of the MSA signal by reducing the number of sequences in the MSA increases contribution from the template and pushes AlphaFold prediction to the conformation specified by a template. This can be easily done by specifying the corresponding parameters of ColabFold, a popular, community-driven front-end to AlphaFold (48).

---

[3] pTM stands for predicted TM-score, and is a measure of AlphaFold's expected global accuracy on a protein or complex. An interface-only version of pTM can be computed for complexes, called ipTM.

## Experimental validation of AlphaFold models

CASP14 provided the initial evidence for AlphaFold's accuracy. However, the evaluation of any computational method is necessarily an ongoing process, involving continued comparison of predictions against new experimental results. This section looks at recent investigations that compare AlphaFold predictions against experimental results on a larger scale, to evaluate different aspects of the method.

### *Single chains*

Molecular replacement, a technique to solve phase problem in macromolecular crystallography, requires a search model that closely resembles the actual contents of the target crystal. Typically this technique works well if the RMSD between model and target is < 1.5Å (89), and it may work at < 2Å RMSD on 50% of atoms (90). Therefore, if AlphaFold predictions can be used for molecular replacement it indicates that they closely resemble the crystal structures. A comprehensive investigation into the use of AlphaFold models for molecular replacement has now been conducted by Terwilliger et al. (23). They took 215 recent PDB structures solved by experimental phasing (indicating that molecular replacement attempts by the original depositors likely failed). They then tried to solve the structures using an AlphaFold search model followed by a fully automated iterative refinement process. Molecular replacement succeeded for 208 / 215 cases, and further automated refinement went on to yield a high quality model for 87% of the structures (at least 50% of C-alpha atoms matching the deposited model to within 2Å).

A second investigation used the same benchmark set of 215 structures to study how closely AlphaFold predictions matched: 1) density maps from the automated refinement process, and 2) deposited PDB models (91). The mean map-model correlation was 0.56 for AlphaFold predictions vs 0.86 for the deposited models. AlphaFold predictions had a median Cα RMSD to the corresponding PDB structure of 1.0 Å. To put this in context, the median RMSD for another PDB structure of the same protein crystallised in a different space group would be 0.6 Å. Confidence scores were predictive of the level of agreement with deposited models, highlighting the importance of referring to these when interpreting predictions. Regions with low confidence (pLDDT < 70) had a median RMSD of 3.5 Å, while for high-confidence regions (pLDDT > 90) the median RMSD was only 0.6 Å. An analysis of side chains indicated that 20% of AlphaFold-predicted side chains are substantially different from the map data and 7% are incompatible with the data; corresponding values for PDB structures in a different space group were 6% and 2%. The authors concluded that AlphaFold predictions are "valuable hypotheses" that "accelerate but do not replace experimental structure determination".

*Complexes*

Evidence for predicted complexes can be obtained at scale using non-structural experimental techniques like cross-linking mass-spectrometry (XL-MS). In this method, chemical cross-linkers are used to covalently fix amino acids that are spatially close. After a protease digestion step, the short cross-linked peptides can be identified by mass spectrometry, providing structural constraints on a protein or protein-protein interface. In one study, in-cell XL-MS was used to search for protein–protein interactions in *Bacillus subtilis*. Based on AlphaFold-Multimer modeling, novel high confidence structures were proposed for 153 dimeric and 14 trimeric protein assemblies. The authors then looked at the cross-link violation rate of AlphaFold-Multimer predictions as a function of the ipTM confidence metric. They found that models with ipTM > 0.85 show especially low rates of cross-link violation and tend to agree with experimental inter-residue distances identified *in situ (58)*. Models in the lower confidence 0.55–0.85 ipTM range showed a wide range of violation rates, and would particularly benefit from checking against independent experimental data.

*Biologically relevant states*

The majority of data used to train AlphaFold comes from crystal structures, and there is a long-standing discussion in the field about how representative these are of proteins in solution and in cells (92).

Cross-linking mass spectrometry is one method that can deliver information about protein states *in situ*. A study by Bartolec et al. generated a high-coverage cross-link dataset for HEK293 cells, then looked at whether these cross links were satisfied in PDB structures as well as in AlphaFold predictions (93). 92% of intra-chain cross links were satisfied in high confidence AlphaFold models for proteins without an experimental structure. This compares favourably with the corresponding cross-link satisfaction rate for PDB structures (89-99%). Another study looked at the 100 best-sampled proteins from intact *Tetrahymena thermophila* cilia, crosslinked *in situ (94)*. AlphaFold models satisfied 86.2% of the experimental cross-links, with 43% of proteins showing no cross-linking violations at all. Observed violations tended to occur in low-confidence regions or between domains. Based on this, the authors concluded that AlphaFold "predicts biologically relevant protein conformations".

Protein structures solved by NMR (nuclear magnetic resonance) are another interesting point of comparison for AlphaFold models. In contrast to crystallography, NMR provides information about *solution-state* protein structures, in the form of spectra and derived coupling constants, e.g. chemical shifts. Fowler and Williamson have reported that AlphaFold outputs are sometimes a better fit to the underlying NMR data than deposited ensembles, based on quality metrics like ANSURR (95).Their investigation looked at 904 human proteins and found that the AlphaFold model had a significantly higher quality score in 30% of cases, while the NMR ensemble was preferred in 2% of cases. Specifically they suggest that AlphaFold can produce more correct hydrogen bonds that persist in solution. A separate focused study on nine small monomeric proteins, absent from the AlphaFold training set, concluded that AlphaFold predictions fit the NMR data almost as well as, or in some cases even better than, experimental structures (96). An investigation of short peptide NMR structures produced more mixed results, with AlphaFold outperforming the other computational methods tested, but showing weaknesses on highly solvated peptides and helix-turn-helix structures (97).

## Conclusions and future directions

The arrival of AlphaFold has been a transformative event in the field of structural biology. We have reviewed some of the many ways the method has been applied in the field, from accelerating experimental structure determination and protein design, to discovering and understanding protein-protein interactions. AlphaFold has also enabled breakthroughs in closely related areas like genetics and molecular biology, for instance allowing development of AlphaMissense, which classifies pathogenicity of the human genetic variants (98). The ability of AlphaFold to scale along with its self-reported confidence metrics have enabled a range of new computational work. We also touched on the ongoing process of evaluating AlphaFold predictions by comparing them against newly deposited structures and experimental data from other methods. A common theme is the importance of interpreting predictions in the light of their confidence metrics, with higher confidence models more likely to prove accurate. In lower confidence bands, a prediction can still provide a useful starting hypothesis, but it is even more important to seek independent experimental data to validate conclusions.

Current AlphaFold limitations can be viewed as an action plan for future development: modelling protein-DNA and protein-RNA complexes, predicting all functional states of a protein, elucidating the effects of point mutations, predicting the binding of ligands / ions, and modelling post-translational modifications; first steps in some of these directions have already been reported (99, 100). Other challenges include predicting larger complexes than ever before, improving predictions for antigen-antibody interactions and orphan proteins, and improving domain positioning for membrane proteins. The inclusion of new categories in CASP15 in 2022 is a sign that the wider field of structure prediction is now moving beyond single protein chains. It also reflects the growing interest in structure prediction and its broader application. We are excited to see what new research future models might enable.

We would like to thank the multitude of people who have contributed to the development and adoption of AlphaFold. This includes not only the team at Google DeepMind, but the many scientists who contributed to the training data, have used AlphaFold in their research, integrated AlphaFold into their tools, expanded its functionality, and provided constructive feedback. The method's widespread adoption and successful use is thanks in large part to the work of this wider community.

Recent progress in structure prediction is a testament to the power of artificial intelligence; we believe this is only the beginning. The future of AI for science is full of promise, and we are only starting to scratch the surface of what is possible.